\documentclass[runningheads]{llncs}
\usepackage[T1]{fontenc}
\usepackage{soul,xcolor}
\usepackage{graphicx}
\usepackage[frozencache,cachedir=.]{minted}
\usepackage{caption}
\usepackage{subcaption}
\usepackage{svg}
\begin{document}
\title{MultiCoSim: A Python-based Multi-Fidelity Co-Simulation Framework}
%
%
\author{Quinn Thibeault\inst{1}\orcidID{0000-0001-8020-9346}
    \and Giulia Pedrielli\inst{1}\orcidID{0000-0001-6726-9790}}
\authorrunning{Q. Thibeault \and G. Pedrielli}
%
\institute{Arizona State University, Tempe, AZ 85281, USA}
\maketitle              
\begin{abstract}
Simulation is a foundational tool for the analysis and testing of cyber-physical systems (CPS), underpinning activities such as algorithm development, runtime monitoring, and system verification. As CPS grow in complexity and scale—particularly in safety-critical and learning-enabled settings—accurate analysis and synthesis increasingly rely on the rapid deployment of simulation experiments. Because CPS inherently integrate hardware, software, and physical processes, simulation platforms must support co-simulation of heterogeneous components at varying levels of fidelity. Despite recent advances in high-fidelity modeling of hardware, firmware, and physics, co-simulation across diverse environments remains challenging. These limitations hinder the development of reusable benchmarks and reduce the accessibility of simulation for automated and comparative evaluation.

Existing simulation tools often rely on rigid configurations, lack automation support, and present obstacles to portability and modularity. Many are configured through static text files or impose constraints on how simulation components are represented and interconnected, making it difficult to flexibly compose systems or integrate components across platforms.

To address these challenges, we introduce MultiCoSim, a Python-based simulation framework that enables users to define, compose, and configure simulation components programmatically. MultiCoSim supports distributed, component-based co-simulation and allows seamless substitution and reconfiguration of components, enabling simulation to be used as a tunable element in system-level design and optimization workflows. We demonstrate the flexibility of MultiCoSim through case studies that include co-simulations involving custom automaton-based controllers, as well as integration with off-the-shelf platforms like the PX4 autopilot for aerial robotics. These examples highlight MultiCoSim's capability to streamline CPS simulation pipelines for research and development.

\end{abstract}
\section{Introduction}

Simulation has become a standard tool for the analysis of cyber-physical systems (CPS), supporting a wide range of activities including algorithm development, runtime monitoring, and verification. Given the complexity and criticality of CPS, accurate analysis and synthesis approaches—particularly for large-scale or safety-critical CPS and learning-enabled CPS (LE-CPS)—require the ability to quickly configure and execute simulation experiments. This is especially important when physical testing is impractical due to cost, safety, or accessibility constraints. Because CPS inherently integrate hardware, software, and physical dynamics, effective simulation platforms must support co-simulation of heterogeneous components across different levels of fidelity, enabling meaningful integration where such composability is appropriate. While significant progress has been made in the development of high-fidelity simulations for hardware, firmware, and physics—including multi-aspect modeling—the ability to coordinate these in heterogeneous co-simulation environments remains limited. This gap poses a major barrier to the creation and dissemination of standardized benchmarks, hindering progress in community-driven evaluation and comparative analysis.

In fact, while a plethora of simulation frameworks are available today, there is little support among them for integrating with automated testing approaches. Many simulation environments are configured using text-based formats, which are difficult to interact with programmatically and have large or complex dependencies that make environment setup and portability difficult. Additionally, executing and managing all of the simulation components is a complex task that can result in fragile execution environments. Furthermore, many simulators also impose restrictions on the representation of simulation components, making it difficult to interface between different simulation environments. As the usage of automated testing approaches increases, addressing these issues becomes paramount to allow users to continue to leverage simulation as part of their testing pipelines.

In this work, we introduce \texttt{MultiCoSim}, a simulation framework that allows users to define, compose, and configure simulator components using Python code. The objective of this library is to support use cases where simulations are composed of several discrete, separately configured components of varying fidelities executing in parallel. The major benefit of this approach is that it becomes much easier to alter or swap components in the simulation depending on the desired simulator performance. This enables optimization-based simulation approaches to treat the simulator configuration as just another system-level variable to optimize during testing. To demonstrate the usefulness of this approach, we provide several case studies combining a variety of self-implemented and off-the-shelf components. These include a rover with a custom automaton-based controller and a quadrotor controlled by the PX4 \cite{px4} autopilot software.

\section{Related Work}
    Many simulator designs and implementations exist targeting different use cases, and in this section, we will discuss some commonly used tools and approaches.

First is MATLAB \cite{matlab}, which is a computational platform that supports a variety of simulation use cases. Most simulator features are organized into toolboxes, some of which are Simulink package for discrete-time finite-element system simulation, and Simscape for electrical power system simulations. However, one of the limitations of MATLAB as a simulation environment is that components implemented using MATLAB may not interface well with other components that are not also implemented in MATLAB, which makes integration with off-the-shelf components difficult. Furthermore, the licensing requirements of MATLAB make it difficult to distribute components to users without a MATLAB license.

Another popular simulator choice is Gazebo \cite{gazebo}, which is a rigid-body simulation engine with a focus on robotics. Gazebo provides several helpful components for defining simulations, including sensor model implementations and a robust transport library for communication and monitoring. However, the configuration of Gazebo simulations is represented using an XML specification called SDFormat \cite{sdformat}, which is difficult to modify using automated tools. Additionally, models built using the Gazebo transport framework are difficult to adapt to other simulators, and off-the-shelf components require a manual effort to integrate as a Gazebo communication node. There are many similar alternatives to Gazebo as well, such as Webots \cite{webots}, CARLA \cite{carla}, and IsaacLab \cite{isaaclab} that have various simulation focuses.

As an alternative to implementing a concrete simulator, some approaches try to define standardized interfaces between components. This design allows simulation components and simulators to be implemented separately and independently for different use cases. One such standard is the Functional Mock-Up Interface \cite{fmi} (FMI), which defines a standard module layout that allows a centralized integrator to advance each component simulation. The benefit of this approach is that functional mock-up units (FMUs) can be generated from a variety of sources, and FMU-based techniques excel when the system is primarily modeled mathematically, as seen in \cite{f-dcs}. However, because the components are all stepped using an integrator, it is difficult to simulate or interface with real-time components like hardware, which limits the types of systems that can be represented.

Another standardization approach is the High-Level Architecture (HLA) \cite{hla}, which creates a system of federated simulation components that also communicate through a well-specified interface. HLA components can be distributed across devices and networks, and are managed by a service called the Run-Time Infrastructure (RTI). This approach can be seen in works such as \cite{dcas,extensible}.

Finally, there are several ad-hoc simulation frameworks and tools that exist to fill specific simulation needs that we will briefly cover. In general, these approaches seek to combine multiple existing tools or libraries, not serve as generic frameworks for simulator composition. Approaches like \cite{kassandra,cps-sim,hybridsim} define their own simulation frameworks by combining multiple tools and providing a unified interface for interacting with them, while approaches like \cite{timetriggered} create simulation environments that compose the outputs of other simulator-generating tools.

\section{Architecture}

\begin{figure}
    \centering
    \includesvg[width=0.9\linewidth]{figures/architecture.svg}
    \caption{Component Diagram of MultiCoSim architecture}
    \label{fig:architecture}
\end{figure}

\subsection{Interfaces}
    The \texttt{MultiCoSim} library is composed of several interfaces, as shown in Figure \ref{fig:architecture}, the most basic being the \textit{Node}, which represents an executing element of the system simulation. Each \textit{Node} is responsible for managing the lifecycle of the simulation component that it represents, and must implement the \textit{Node.stop} method to shut down the component when the simulation is complete. Extending the base \textit{Node} interface is the \textit{CommunicationNode}, which defines a simulation component that can receive and send messages using the \textit{CommunicationNode.send} method. This enables users to communicate with the \textit{Node} at run-time, providing system input or configuration, and receiving back system state or status information.

    Nodes are managed by instances of the \textit{Simulation} interface. The \textit{Simulation} is responsible for shutting down all of its member components using the \textit{Simulation.stop} method, and for providing access to the individual nodes through the \textit{Simulation.get} method. \textit{Node} instances are uniquely referenced using a data type called the \textit{NodeId}, which is generated when the \textit{Component} is added to the \textit{Simulator}.

    The creation of \textit{Nodes} is handled by instances of the \textit{Component} interface. A \textit{Component} contains all of the configuration necessary to execute the simulation element it represents, and is responsible for producing the \textit{Node} instance that is stored in the \textit{Simulation}. Creation of the \textit{Node} is done by calling the \textit{Component.start} method, which is provided a single parameter representing the execution environment of the simulation component and is responsible for starting the execution of the component before returning the \textit{Node} instance. 

    The final interface provided by \texttt{MultiCoSim} is the \textit{Simulator}, which represents a set of configured simulator \textit{Components}. Each \textit{Simulator} implementation is responsible for initializing the execution environment, executing each of its components using the environment, and aggregating them into a \textit{Simulation} value. \textit{Components} can be added to the simulation using the \textit{Simulator.add} method, which should return a \textit{NodeId} value uniquely identifying the component so that it can be retrieved later from the \textit{Simulation}.

    The decision to represent the core functionality of the library as interfaces imposes very few restrictions on the implementations provided by the user. This allows the library to remain very general, and for the implementations described in the next section to be reused in a variety of simulator configurations. Furthermore, by not imposing as many restrictions on the component implementations, we can support potentially heterogeneous simulator environments with components of multiple different types, run-times, or technologies.

\subsection{Implementation}

    \texttt{MultiCoSim} provides several interface implementations that are defined using Docker containers. The most primitive of these implementations is the \textit{ContainerNode}, which implements the \textit{Node} interface and represents a single simulation component executing inside a container. Constructing a \textit{ContainerNode} requires providing an executing container, which will then be managed by the node during the simulation lifecycle. Corresponding to the \textit{ContainerNode} is the \textit{ContainerComponent}, which implements the \textit{Component} interface and is responsible for constructing the container that is managed by the node. Creating a \textit{ContainerComponent} requires a container image, as well as a command that will executed inside the container. The \textit{ContainerComponent} also must be provided an existing Docker network as part of the execution environment. The container constructed by this component will be attached to this network so that it can communicate with the other \textit{ContainerNodes} that are part of the simulation. Managing the orchestration of the \textit{ContainerComponents} is the \textit{ContainerSimulator}, an implementation of the \textit{Simulator} interface that creates the execution environment provided to each \textit{ContainerComponent} and manages the collection of components that will be part of the simulation. When started, the \textit{ContainerSimulator} creates a \textit{ContainerSimulation} that implements the \textit{Simulation} interface and manages all of the simulation nodes.

    \texttt{MultiCoSim} also provides more advanced interface implementations for different use-cases. The first is the \textit{AttachedComponent}, which represents a group of containers which share the same network stack. The difference between a \textit{ContainerComponent} and an \textit{AttachedComponent} is that, instead of each container in the attached component being connected to the provided Docker network, only the host container of the \textit{AttachedComponent} is connected to the network and each other container of the component is configured to re-use the host's network stack. The advantage of this is that all of the components sharing the network stack can communicate with each other via the \textit{localhost} interface, rather than the network interface, which can help solve problems with container communication. Another advanced component implementation is the \textit{FirmwareComponent}, which allows for bi-directional communication with the simulation running in the docker container by implementing the \textit{CommunicationNode} interface. Communication is accomplished by binding a port in the container to a random port on the host, and using $\emptyset$-MQ sockets to transport the messages.

    The implementations provided by this library are fully type-annotated in accordance with PEP-484 \cite{pep484}. Therefore, correct usage of these components can also be enforced using a type checker such as PyRight or MyPy. When using a type checker, some classes will carry additional type information making them easier to use. For example, the \textit{NodeId} class carries the information about the \textit{Node} type that is will create, thus editing tools will be able to provide richer suggestions regarding the methods available once the node is retreived from the simulation.

\subsection{Integration with existing tools}
    \texttt{MultiCoSim} currently provides integrations with two off-the-shelf tools: the PX4 flight control software, and the Gazebo rigid-body simulator. These integrations include a \textit{Component} implementation for managing the execution of the tool, and a Docker image that is responsible for executing the tool. The only \textit{Component} implementations used to define these integrations are the ones described above, thus these integrations also serve as examples for constructing integrations for other tools.
    
    
    Access to the Gazebo simulator is provided by the \textit{GazeboContainerComponent} class, which is defined as a wrapper around the \textit{ContainerComponent} class. This component allows the user to configure the physics backend of the simulation, as well as modify model files to remap sensor topics. This functionality comes from a program embedded in the provided Gazebo Docker image that accepts a template file defining the simulation world, and produces a new file with the updated configuration that is then used to execute the Gazebo simulation. The configuration options provided to the component are translated into command-line options that are passed to the embedded program. The Gazebo image is designed to be extended, as in the case of the PX4 where the PX4-specific models are also embedded within the image.

    Interfacing with the PX4 \cite{px4} flight control software can be done through either the \textit{PX4Component} class or the \textit{PX4} class. The \textit{PX4Component} is a \textit{ContainerComponent} wrapper that is specifically configured to launch the provided PX4 Docker image and can be added to any \textit{ContainerSimulator} instance. Conversely, the \textit{PX4} class is a \textit{Simulator} implementation that wraps around a \textit{ContainerSimulator} and adds a a \textit{PX4Component} to the simulation upon initialization. Both of these classes require a Gazebo configuration object to be provided because the PX4 requires Gazebo for simulation. After simulation has been started, either the \textit{NodeId} returned from adding a \textit{PX4Component} to the simulator, or the \textit{PX4Simulation.firmware} attribute can be used to access the PX4 firmware component and then a mission can be executing by sending the runtime configuration as a message using the \textit{PX4Node.send} method. An example using the PX4 components provided by \texttt{MultiCoSim} can be seen in Listing~\ref{lst:multicosim-px4}.

    %
    %
    %

\section{Case Studies}
    To demonstrate the usefulness of our library, we provide several case studies demonstrating different use cases. Our first example , shown in Figure \ref{lst:multicosim-px4}, is an unmodified PX4 simulation. In this example, we initialize a PX4 simulator, which requires a Gazebo configuration.  In this case, we can use the \textit{PX4} helper class, which ensures that a Gazebo component is also created and managed to avoid any runtime errors. We are then able to start the simulation and send the PX4 a mission to execute. This use case is helpful for users who are interested in simulating PX4 missions without any additional components, but who may be interested in customizing either the mission or the simulation parameters. 


\begin{figure}
    \centering
    \begin{minipage}[t]{0.48\textwidth}
        \begin{minted}[linenos,autogobble,python3,xleftmargin=0.1\textwidth,fontsize=\scriptsize]{python}
            import multicosim.px4 as px4
    
            gazebo = px4.Gazebo()
            simulator = px4.PX4(gazebo)
            simulation = simulator.start()
            mission = [
                px4.Waypoint(
                    lat=0.025,
                    lon=-0.018,
                    alt=25.0,
                )
            ]
            result = simulation.firmware.send(
                px4.Configuration(mission)
            )
        \end{minted}
        \captionof{listing}{PX4 Simulation Example}
        \label{lst:multicosim-px4}
    \end{minipage}
    \hfill
    \begin{minipage}[t]{0.5\textwidth}
        \begin{minted}[linenos,autogobble,xleftmargin=0.1\textwidth,fontsize=\scriptsize]{python}
            import multicosim as mcs

            REPO = "ghcr.io/cpslab-asu"
            FW_IMG = f"{REPO}/examples/rover/controller"
            GZ_IMG = f"{REPO}/examples/rover/gazebo"

            class Mission:
                speed: int

            port = 5556
            rover = mcs.FirmwareComponent(
                image=FW_IMG,
                command="controller --port {port}",
                port=port,
            )
            gazebo = mcs.GazeboContainerComponent(
                image=GZ_IMG,
            )

            simulator = mcs.ContainerSimulator()
            simulator.add(gazebo)
            
            node_id = simulator.add(rover)
            simulation = simulator.start()
            node = simulation.get(node_id)
            trajectory = node.send(Mission(speed=5))
        \end{minted}
        \captionof{listing}{Rover Simulation Example}
        \label{lst:multicosim-rover}
    \end{minipage}
\end{figure}

For our next case study, we extend our simple example to include a custom component, which in this case is a noisy GPS component. In this scenario the actual GPS sensor simulated by Gazebo is injected with additional noise by an external program before the data is read by the PX4 firmware. This is accomplished by changing the Gazebo transport topic that the sensor uses to publish its data from the topic the firmware expects to a different topic, and then subscribing to the sensor topic with the noise generator to receive the sensor messages, and finally broadcasting the modified sensor messages on the topic expected by the firmware. This scenario can be helpful to users who may be interested in simulating cyber-physical attacks on the system, as the additional components can represent attackers in addition to custom sensor implementations.

As another example, we demonstrate using a custom controller implementation instead of the PX4 for actuating an Ackermann steering model rover. In this example, which is shown in Figure \ref{lst:multicosim-rover}, we have implemented the controller as a finite state automaton (FSA) that interacts with the Gazebo model over the transport layer. Unlike the PX4 simulator, this simulator has one less layer, which is because we do not implement a \textit{Node} like the \textit{PX4Node} which manages the execution of the \textit{FirmwareComponent} and \textit{GazeboComponent} sub-components. In this case, the user is responsible for managing all the components and ensuring that all required components are added. We are also required to create an implementation-specific Gazebo container with the Gazebo model we are interested in simulating. This scenario may be useful for testing custom systems instead of using already existing options.

Finally, our last example, shown in Listing \ref{listing:psytaliro-example}, demonstrates how simulator configuration can be integrated into search-based test generation approaches using the $\Psi$-TaLiRo framework. In this use case, the simulation configuration is treated as a set of system-level variables that are included in the input space, allowing the optimizer to consider both the system inputs and the simulation parameters. In the context of system verification, this could be used to find the lowest fidelity of a simulator that still produces a falsification, or help improve performance by avoiding higher-costed simulations until absolutely necessary. Due to the extensible nature of the \texttt{MultiCoSim} framework, integration with a search-based test generator like $\Psi$-TaLiRo could also be expanded to encompass the simulation components themselves in addition to the component configurations, allowing for switching component implementations.

\section{Conclusion}
    In this paper, we have introduced \texttt{MultiCoSim}, a Python-based framework for multi-fidelity co-simulation. This framework is based on several extensible interfaces, which allows users to implement their own components and simulations, and also provides several helpful implementations to assist users in getting started quickly. We demonstrate the usefulness of this framework using several case studies of varying complexity, and provide insight into integrating this toolbox with search-based test generation. The \texttt{MultiCoSim} framework is available to use on GitHub at the following URL: \url{https://github.com/cpslab-asu/multicosim}.

In the future, we would like to increase the number of implementations provided by this library, eventually supporting other off-the-shelf controllers like ArduPilot and CARLA. We would also like to continue to explore new component interfaces that can be generalized to support even more use-cases. Finally, we would like to provide ready-made integrations with $\Psi$-TaLiRo to accommodate the more common use cases that we encounter during our own research.

\begin{credits}

\end{credits}
%
%
%
\nocite{*}
\bibliographystyle{splncs04}
\bibliography{references}
%
%
%
%
%

\newpage
\appendix
\section{$\Psi$-TaLiro Example}
\begin{listing}[h!]
    \centering
    \begin{minted}[linenos,autogobble,xleftmargin=0.1\textwidth]{python}
        import multicosim as mcs
        import multicosim.gazebo as gz
        import multicosim.px4 as px4
        import staliro
        import stalio.models
        import staliro.optimizers as opts
        import staliro.requirements.rtamt as req

        @staliro.models.model
        def mymodel(sample: staliro.Sample) -> staliro.Trace:
            gazebo = px4.Gazebo(
                backend=gz.ODE(iterations=round(sample.static["iters"])),
                step_size=sample.static["step_size"],
            )
                
            system = px4.PX4(gazebo)
            simulation = system.start()
            mission = [
                px4.Waypoint(
                    sample.static["lat"],
                    sample.static["lon"],
                    sample.static["alt"],
                )
            ]
            trajectory = simulation.firmware.send(px4.Configuration(mission))

        spec = req.parse_dense("always alt > 0.0")
        opt = opts.UniformRandom()
        options = staliro.TestOptions(
            static_inputs = {
                "iters": (1, 100),
                "step_size": (0.001, 0.01),
                "lat": (0, 180.0),
                "lon": (-180.0, 180.0),
                "alt": (25.0, 50.0),
            },
            iterations=50
        )
        res = staliro.test(mymodel, spec, opt, options)
    \end{minted}
    \caption{Example $\Psi$-TaLiRo program integrating MultiCoSim}
    \label{listing:psytaliro-example}
\end{listing}

\end{document}